\newcommand*{\ATLASLATEXPATH}{latex/}
\documentclass[11pt,a4paper,UKenglish,texlive=2016]{article}
\pdfoutput=1
\usepackage{journal/jinstpub}

\usepackage[biblatex=false]{\ATLASLATEXPATH atlaspackage}

\graphicspath{{logos/}{figures/}}

\hypersetup{pdftitle={ATLAS draft},pdfauthor={The ATLAS Collaboration}}

\usepackage{multirow}
\DeclareSIUnit{\sample}{S}
\title{Performance in beam tests of Carbon-enriched irradiated Low Gain Avalanche Detectors for the ATLAS High Granularity Timing Detector}

\makeatletter
\newcommand\footnoteref[1]{\protected@xdef\@thefnmark{\ref{#1}}\@footnotemark}
\makeatother

\author[a]{S.~Ali,}
\author[b]{H.~Arnold,}
\author[c]{S. L.~Auwens,}
\author[d]{L. A.~Beresford,}
\author[e]{D. E.~Boumediene,\footnote{\label{note1}Corresponding author.}}
\author[e]{A. M.~Burger,}
\author[f]{L.~Cadamuro,}
\author[g]{L.~Castillo Garc\'{i}a,\footnoteref{note1}}
\author[e]{L. D.~Corpe,}
\author[h]{M. J.~Da Cunha Sargedas de Sousa,}
\author[i]{D.~Dannheim,}
\author[i]{V.~Dao,}
\author[i]{A.~Gabrielli,}
\author[j]{Y.~El Ghazali,}
\author[k]{H.~El Jarrari,}
\author[g]{V.~Gautam,}
\author[g,l]{S.~Grinstein,}
\author[m]{J.~Guimar\~{a}es da Costa,}
\author[i]{S.~Guindon,}
\author[m]{X.~Jia,}
\author[n]{G.~Kramberger,}
\author[h]{Y.~Liu,}
\author[h]{K.~Ma,}
\author[f]{N.~Makovec,}
\author[i]{S.~Manzoni,}
\author[o]{I.~Nikolic,}
\author[e]{O.~Perrin,}
\author[o]{V.~Raskina,}
\author[p]{M.~Robles Manzano,}
\author[i]{A.~Rummler,}
\author[k]{Y.~Tayalati,}
\author[o]{S.~Trincaz-Duvoid,\footnoteref{note1}}
\author[b]{A.~Visibile,}
\author[m]{S.~Xin,}
\author[h]{L.~Xu,}
\author[h]{X.~Yang,}
\author[h]{and X.~Zheng.}

\affiliation[a]{Academia Sinica, 128, Section 2, Academia Road, Nangang District, Taipei City, Taiwan 115}
\affiliation[b]{Nikhef National Institute for Subatomic Physics and University of Amsterdam, Postbus 41882
NL - 1009 DB Amsterdam, Netherlands}
\affiliation[c]{Institute for Mathematics, Astrophysics and Particle Physics, Radboud University/Nikhef, Nijmegen, P.O. Box 9010, 6500 GL Nijmegen, Netherlands}
\affiliation[d]{Deutsches Elektronen-Synchrotron (DESY), Notkestraße 85, 22607 Hamburg, Germany}
\affiliation[e]{Laboratoire de Physique de Clermont-Ferrand (LPC), Universite Clermont Auvergne, Campus Universitaire des Cézeaux, 4 Avenue Blaise Pascal, 63178 Aubière Cedex, France}
\affiliation[f]{Laboratoire de Physique des 2 Infinis Irène Joliot Curie (IJCLab), 15 Rue Georges Clemenceau, 91400 Orsay, France}
\affiliation[g]{Institut de F\'{i}sica d'Altes Energies (IFAE), The Barcelona Institute of Science and Technology (BIST), \\Carrer Can Magrans s/n, Edifici Cn, Campus UAB, E-08193 Bellaterra (Barcelona), Spain}
\affiliation[h]{Department of Modern Physics and State Key Laboratory of Particle Detection and Electronics, University of Science and Technology of China (USTC), 96 JinZhai Road Baohe District, Hefei, Anhui, 230026, China}
\affiliation[i]{Conseil Europ\'{e}en pour la Recherche Nucl\'{e}aire (CERN), Esplanade des Particules 1, CH-1211 Meyrin, Switzerland}
\affiliation[j]{Facult\'{e} des Sciences, Universit\'{e} Ibn-Tofail, Avenue de l'Universit\'{e}, K\'{e}nitra, Morocco}
\affiliation[k]{Universit\'{e} Mohammed V de Rabat, Avenue des Nations Unies, Agdal, Rabat, Morocco}
\affiliation[l]{Instituci\'{o} Catalana de Recerca i Estudis Avan{\c c}ats (ICREA), Passeig de Llu\'{i}s Companys, 23, 08010 Barcelona, Spain}
\affiliation[m]{Institute of High Energy Physics (IHEP), 19 Yuquan Road, Shijingshan District, Beijing, China}
\affiliation[n]{Jo\v{z}ef Stefan Institute (JSI), Jamova cesta 39, 1000 Ljubljana, Slovenia}
\affiliation[o]{Laboratoire de Physique Nucl\'{e}aire et de Hautes Energies (LPNHE), Sorbonne Universit\'{e}, Universit\'{e} de Paris, CNRS/IN2P3, Paris, France}
\affiliation[p]{Johannes Gutenberg Universit$\ddot{a}$t Mainz, Saarstra{\ss}e 21, 55122 Mainz, Germany}

\emailAdd{djamel.boumediene@cern.ch, lucia.castillo.garcia@cern.ch, trincaz@lpnhe.in2p3.fr}

\abstract{
The High Granularity Timing Detector (HGTD) will be installed in the ATLAS experiment to mitigate pile-up effects during the High Luminosity (HL) phase of the Large Hadron Collider (LHC) at CERN. Low Gain Avalanche Detectors (LGADs) will provide high-precision measurements of the time of arrival of particles at the HGTD, improving the particle-vertex assignment. To cope with the high-radiation environment, LGADs have been optimized by adding carbon in the gain layer, thus reducing the acceptor removal rate after irradiation. 
Performances of several carbon-enriched LGAD sensors from different vendors, and irradiated with high fluences of 1.5~and~$2.5\times10^{15}$~n$_{eq}$/cm$^{2}$, have been measured in beam test campaigns during the years 2021 and 2022 at CERN SPS and DESY. This paper presents the results obtained with data recorded by an oscilloscope synchronized with a beam telescope which provides particle position information within a resolution of a few $\mu$m. Collected charge, time resolution and hit efficiency measurements are presented. In addition, the efficiency uniformity is also studied as a function of the position of the incident particle inside the sensor pad.
}


\arxivnumber{2303.07728}

\begin{document}

\maketitle

\section{Introduction}
\label{sec:intro}
The High Granularity Timing Detector (HGTD)~\cite{hgtd_tp,hgtd_tdr} will be added to the ATLAS experiment~\cite{atlas} during the so-called Phase-II upgrade preceding the high-luminosity phase of the LHC (HL-LHC)~\cite{hllhc,hllhc_tdr}. Low Gain Avalanche Detector (LGAD) sensors~\cite{lgad} have been extensively studied during the R\&D phase of the HGTD project. These sensors must meet performance targets regarding time resolution of \SI{50}{\pico\second} (\SI{70}{\pico\second}) per hit, collected charge $>$\SI{4}{\femto\coulomb} and hit efficiencies of 97\% (95\%) at the start (end) of their lifetime. The expected maximum fluence is $2.5 \times 10^{15}$~n$_{eq}$/cm$^{2}$~\cite{hgtd_tdr}. These requirements should be achieved taking into account a discriminator of about \SI{2}{\femto\coulomb} of the front-end ASIC foreseen for HGTD, ALTIROC~\cite{altiroc0}.

Many previous beam test campaigns were devoted to quantifying LGAD performances before~\cite{hgtd_tb_paper,altiroc0_results} and after~\cite{irrad_lgad_tb,hgtd_tb_paper_2018} irradiation. A variety of pad structures, such as single-pad diodes and segmented arrays of pad diodes with various granularities, from several vendors were tested to prove they meet the requirements to be operated over the entire HL-LHC lifetime. Irradiated devices met the required performance up to $10^{15}$~n$_{eq}$/cm$^{2}$, however, some optimization was still required.

In beam tests~\cite{hgtd_tb_paper_2018} as well as in laboratory measurements~\cite{instruments2022}, it was observed that the addition of carbon in the gain layer helps reducing the operating voltage needed to collect the same charge by a reduction of the acceptor removal rate after irradiation, hence improving the radiation hardness~\cite{ferrero} of the sensors. This is extremely important at higher fluences where standard LGADs need a rather high voltage to maintain their performance (gain, time resolution) after irradiations.

Beam test campaigns conducted in the last three years have been devoted to study the performances of LGAD sensors from different vendors and irradiated with higher fluences, $1.5-2.5 \times 10^{15}$~n$_{eq}$/cm$^{2}$. Some of these beam tests were also used to analyze the sensor mortality rates when exposed to high-intensity beams to mimic end-of-life conditions of HGTD LGADs at the HL-LHC. A second paper is under preparation and will describe this particular phenomenon. 

The present paper aims to describe the performance of a selection of promising LGAD sensors with carbon enriched gain layer that could equip the future HGTD. The tested LGADs are described in Section~\ref{sec:sec_sensors}. The response of the LGADs to particles has been tested in test beam facilities with an experimental set-up presented in Section~\ref{sec:sec_tbsetup}. The collected data are analyzed according to the method described in Section~\ref{sec:sec_analysis}. The results of these tests are presented in Section~\ref{sec:sec_results}.

\section{Sensors}
\label{sec:sec_sensors}

The prototypes tested in this paper were manufactured by Fondazione Bruno Kessler (FBK) in Povo, Italy and by the Institute of Microelectronics (IME) of the Chinese Academy of Sciences in China, where the latter are of two different designs, one by the Institute of High Energy Physics (IHEP) and the other by the University of Science and Technology of China (USTC), both in China. An additional device produced by Centro National de Microelectr\'{o}nica (CNM) Barcelona, Spain is included in the tests for timing purposes due to its excellent time resolution before irradiation. However, its full performance is not described in this paper as it was studied in the past~\cite{hgtd_tb_paper}. 

In order to study the LGAD performance after irradiation, the sensors were exposed to fluences up to 2.5$\times$10$^{15}$~n$_{eq}$/cm$^{2}$ at the TRIGA reactor in Ljubljana, Slovenia with fast neutrons. Table~\ref{tab:sensors} lists the LGAD sensors measured in the beam tests, including the vendor, the sensor IDs, the implant of the multiplication layer as well as the irradiation type and fluence. It also includes the device name assigned to each sensor for easier reference in the text: a concatenation of the sensor vendor (CNM, FBK, USTC, IHEP) and the irradiation level in units of 10$^{15}$~n$_{eq}$/cm$^{2}$. For instance, CNM-0 refers to an unirradiated CNM LGAD, which was used in all beam tests as a time reference.

\begin{table}[h]
\centering
\caption{List of CNM, FBK, USTC-IME and IHEP-IME LGAD sensors studied in the 2021 and 2022 beam test periods including the information on the implant of the multiplication layer, the irradiation level and type and the facility where they were tested.}
{\footnotesize
\begin{tabular}{|c|c|c|c|c|c|c|}
\hline
\textbf{Device name} & \textbf{Vendor} & \textbf{Sensor ID} & \textbf{Implant} & \textbf{Irradiation type} & \begin{tabular}{@{}c@{}}\textbf{Fluence} \\ \textbf{[n$_{eq}$/cm$^{2}$]}\end{tabular} & \textbf{Tested at} \\
\hline
CNM-0 & CNM & W9LGA35 & boron & unirradiated & -- & DESY/CERN \\
\hline
FBK-1.5 & FBK & UFSD3.2 W19 & boron + carbon & neutrons & 1.5$\times$10$^{15}$ & DESY/CERN \\
FBK-2.5 & FBK & UFSD3.2 W19 & boron + carbon & neutrons & 2.5$\times$10$^{15}$ & DESY/CERN \\
\hline
USTC-1.5 & USTC-IME & v2.1 W17 & boron + carbon & neutrons & 1.5$\times$10$^{15}$ & DESY \\    
USTC-2.5 & USTC-IME & v2.1 W17 & boron + carbon & neutrons & 2.5$\times$10$^{15}$ & DESY \\
\hline
IHEP-1.5 & IHEP-IME & v2 W7 Q2 & boron + carbon & neutrons & 1.5$\times$10$^{15}$ & DESY/CERN \\ 
IHEP-2.5 & IHEP-IME & v2 W7 Q2 & boron + carbon & neutrons & 2.5$\times$10$^{15}$ & CERN \\
\hline 
\end{tabular}
}
\label{tab:sensors}
\end{table}

\begin{figure}[h]
	\centering 
    \includegraphics[width=0.5\textwidth]{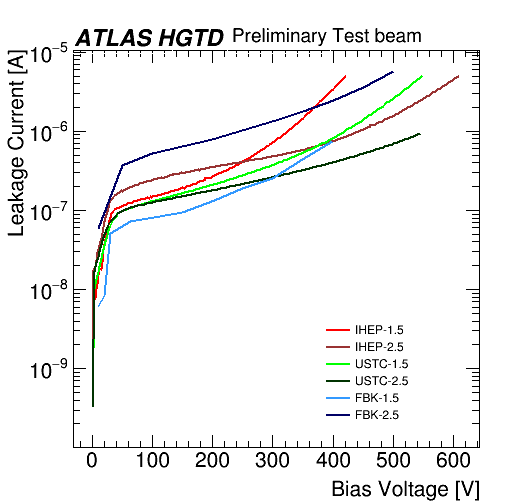}
    \caption{Leakage current-voltage dependence for all sensors performed at \SI{-30}{\degree C}.}
    \label{fig:iv}
\end{figure}

Figure~\ref{fig:iv} shows the leakage current-voltage (I-V) characteristics for several tested sensors. The I-V measurements for all sensors revealed a leakage current below \SI{5}{\micro\ampere} at about \SI{500}{\V}. 

\subsection{FBK devices}
These samples are LGAD sensors optimized for timing, also known as Ultra-Fast Silicon Detectors (UFSDs)~\cite{nicolo,vsola,4Dtracking} which combine an optimized sensor geometry with a moderate internal gain to achieve a time resolution of \SIrange{30}{40}{\pico\second} for minimum ionizing particles (MIPs).

This production run is called UFSD3.2 and aimed to investigate different gain layer designs where two different gain depths were considered: shallow and deep. The gain layer depth is \SI{2}{\micro\metre} and the active thickness in most of the wafers is \SI{45}{\micro\metre}. In particular, with respect to the previous production run UFSD3, this one has an optimization of the carbon level and a different array inter-pad gap. Devices in W19 have a thinner bulk for a better time resolution and a combination of deep gain layer and carbon implantation. Different boron doses are considered: 0.70, 0.74, 0.78 (W19), 0.94 and 0.98 (W7) in arbitrary units (a.u.). as well as carbon doses: 0.4, 0.6 (W19), 0.8 and 1 (W7) in a.u. Different diffusion cycles are used, in particular, low (W7) and high (W19) implantation energies. The units could not be made available due to non-disclosure agreement and are not the same for all the producers mentioned in this paper.
%

The combination of deep gain layer, high doping and carbon implantation shows exceptional performance for the devices from wafer W19 compared with those from wafer W7 (shallow gain layer, carbon, same type as FBK old production UFSD3, \SI{55}{\micro\metre} active thickness). W19 shows improved behavior: lower voltage for a similar collected charge, better time resolution and lower power dissipation. The carbon dose for devices in W19 is 0.6 a.u. The acceptor removal mechanism reduces the effective doping concentration in the gain layer and the sensor performance can be partially recovered by increasing the applied bias voltage after irradiation. The gain layer depletion voltage, V$_{gl}$, varies with the fluence as given by:
\begin{equation}
V_{gl}=V_{gl0} \times exp(-c \times \phi_{eq})
\end{equation}
where c is the acceptor removal coefficient. For these devices (W19), this coefficient is $1.73 \times 10^{-16}$~cm$^{2}$~\cite{tuninggaincarbon}.

\subsection{IHEP-IME devices}
The production run for the IHEP-IME devices considered in this paper is called version 2~\cite{ihepsensor}, yielding silicon wafers with a \SI{50}{\micro\metre}-thick high-resistivity epitaxial layer, from which 3 out of 8 are carbonated. The wafers were split into four quadrants. The implantation and diffusion scheme is carbon implantation and diffusion plus boron implantation and diffusion (CHBL) for wafers W7 and W8, and CLBL for wafer W4. The carbon doses differed from quadrant to quadrant in the wafer and the values that were considered in the full production are 0.2, 0.5, 1, 3, 5, 6, 8, 10 and 20 a.u. The most promising devices belong to wafer W7 and quadrant Q2 with a carbon dose of 0.5 a.u. The acceptor removal coefficient for these devices (W7Q2) is 1.14$\times$10$^{-16}$~cm$^{2}$. The breakdown voltage, V$_{bd}$, is around \SI{170}{\volt}. Before irradiation V$_{gl}$ is \SI{24.4}{\volt}.

\subsection{USTC-IME devices}
The production run for the USTC-IME devices considered in this paper is called version 2.1~\cite{ustcsensor}, yielding 8-inch wafers with a \SI{50}{\micro\metre} epitaxial layer from which 4 out of 5 are carbonated. The gain layer dose is medium and the energy is low. The implantation is boron and the carbon dose is 1 a.u. The wafers include small arrays. The average breakdown voltage is about \SI{190}{\volt}, the V$_{bd}$ is in the range of \SIrange{150}{240}{\volt}. An under bump metallization was deposited on the wafers and afterwards were diced. This version aimed to lower the V$_{bd}$ to enhance the irradiation hardness further and improve and evaluate the yield of the 15$\times$15 arrays. Devices from wafer W17 show impressive radiation hardness. The acceptor removal coefficient for these devices (W17) is about 1.23$\times10^{-16}$~cm$^2$. 

Table~\ref{tab:sensorsprop} summarizes the different properties of the tested sensors in terms of V$_{gl0}$, diffusion scheme and acceptor removal coefficient.

\begin{table}[h]
\centering
\caption{Summary of properties for FBK, USTC-IME and IHEP-IME LGAD sensors studied in the 2021 and 2022 beam test periods.}
{\footnotesize
\begin{tabular}{|c|c|c|c|}
\hline
\textbf{Device name} & \textbf{V$_{gl0}$ [V]} & \textbf{Diffusion} & \textbf{c [cm$^{2}$]} \\
\hline
FBK-1.5/2.5 & 50 & H & $1.73 \times 10^{-16}$ \\
\hline
USTC-1.5/2.5 & 27 & L & $1.23 \times 10^{-16}$ \\    
\hline
IHEP-1.5/2.5 & 25 & CHBL & $1.14 \times 10^{-16}$ \\ 
\hline 
\end{tabular}
}
\label{tab:sensorsprop}
\end{table}



\section{Test beam set-up}
\label{sec:sec_tbsetup}
One HGTD beam test campaign was conducted at DESY~\cite{desyfacility} on beamline 24 using a \SI{5}{\giga\electronvolt} electron beam followed by a campaign at the CERN SPS~\cite{spsfacility} H6A line using a high-momentum \SI{120}{\giga\electronvolt} pion beam. The beam profile size was about $2 \times 2$~cm$^{2}$ in both cases. The set-up used at DESY was identical to the set-up described in detail in~\cite{hgtd_tb_paper_2018}.

The LGADs were mounted on custom readout boards~\cite{ucsc_readoutboard} with on-board and external amplification stages to enhance their signal. The position of the devices under test (DUTs) was coherently controlled by a micrometric $x-y$ motor stage in the plane perpendicular to the beam axis ($z$ direction). 

The set-up included a beam telescope in which six planes were used to track the incident charged particles. The telescope's role was to provide the position of the incoming particles in the frame of each DUT to perform efficiency measurements.
A four-channel oscilloscope was used to sample the waveforms from the DUTs. A trigger system was used to initiate the data acquisition (DAQ) of the oscilloscope and the tracking planes, which can be uniquely linked by a common event number. The trigger fired on a signal from the tracking system, except during oscilloscope dead time or times during which the telescope or reference tracking plane was in a ``busy'' state.

\subsection{Set-up at DESY}
\label{sec:DESYsetup}
The time resolution of each DUT was estimated by testing it with two other devices, including a time reference system. At DESY, a silicon photomultiplier (SiPM) combined with a quartz bar provided a time reference that was acquired by the DAQ together with two DUTs. The measured time resolution from this device was found to be 62.6$\pm$0.6~ps 
for an operating voltage of \SI{27}{\V}. This time resolution was degraded compared to the previous measurements~\cite{hgtd_tb_paper_2018} due to ageing and irradiation effects.

The set-up included a EUDET telescope based on six MIMOSA~\cite{telescope} pixel planes. It was combined with an FE-I4~\cite{fei4} readout chip-based module to reconstruct the tracks. A plastic scintillator coupled to a photomultiplier tube was also used in the front of the set-up. The coincidence between the FE-I4 hitOR signal and the scintillator signal was used to trigger the data-taking. The FE-I4 chip was configured to consider only signals in a region of interest (ROI) compatible with the DUT geometries. The trigger logic was handled by a programmable Trigger Logic Unit (TLU)~\cite{eudettlu}. Moreover, a NIM logic circuit was implemented to generate a busy signal covering the time the oscilloscope needed to read out its data buffer.
Dry ice-based cooling was used to maintain, but not control, the temperature of the DUTs between \SI{-43}{\degree C} to \SI{-25}{\degree C}. The monitoring of the temperature and humidity was performed with Pt100 probes.  

\subsection{Set-up at CERN}
\label{sec:CERNsetup}
The set-up at CERN relied on the MALTA telescope~\cite{MALTA}. MALTA is a full-scale monolithic pixel detector implemented in TowerJazz \SI{180}{\nano\metre} CMOS technology. It contains a small pixel electrode that allows for the implementation of a fast, low-noise and low-power front-end, which is sensitive to the charge released by ionising radiation in a 20–25~$\mu$m deeply depleted region. The novel asynchronous matrix architecture is designed to ensure low power consumption and high rate capability.

The MALTA telescope installed in the North Area of CERN, on the H6 beamline, consists of six MALTA chips, three on each side of the DUT. The timing reference is a second LGAD sensor, CNM-0, which has been placed next to the cooling box. It has been calibrated in the lab and in beam tests to a time resolution of 54.8~ps at room temperature. Its time resolution at a temperature of \SI{-30}{\degree C} is \SI{35}{\pico\second}. For triggering on a particle, the second MALTA sensor is used in coincidence with a scintillator placed at the back of the telescope. If the two sensors record a signal, an FPGA-based triggering system records the telescope data from all six planes and the waveforms from both the DUT and the LGAD used as a timing reference. A remote-controlled cooling system was used to maintain the temperature of the DUTs at  \SI{-20}{\degree C} in a hermetic box.


\section{Data analysis}
\label{sec:sec_analysis}
Two independent systems were used to collect the data processed in this analysis: the oscilloscope, which provides data on the LGAD and SiPM waveforms, and the telescope (MALTA at CERN and EUDET-type at DESY) which gives information on the particle tracks.  
 
This section describes the general methodology used to reconstruct and process the information from these two chains, and how it is used to derive the physical quantities for the analysis.  

\subsection{Track reconstruction for data taken during DESY beam tests with EUDET-type telescope}
\label{sec:tracking_Patrack}

The EUDET-type telescope and FE-I4 together provided tracking information allowing the reconstruction of the trajectory of particles and the identification of the specific position where the DUT was hit.

The tracking capability of the EUDET-type telescope was provided by six MIMOSA planes as explained in section~\ref{sec:DESYsetup}. 
The positions of the MIMOSA, FE-I4 and DUT planes were recorded with a precision of \SI{1}{\milli\metre} in the $z$-direction along the beam line. The positions of the hits from each MIMOSA plane, together with their respective $z$-coordinate, were used to reconstruct particle trajectories and the ($x$,$y$)-coordinates of hits in the DUT planes. 

After the removal of so-called ``hot'' pixels from the MIMOSA planes, identified as those with an occupancy higher than ten times the average, the remaining hits were grouped into clusters. Only clusters with a maximum of 20 hits were used for tracking. The cluster coordinates were the centroid of the hit coordinates in $x$ and $y$. Only events with exactly one cluster in the FE-I4 were considered. The MIMOSA planes were aligned by iteratively shifting the planes' coordinates in $x$ and $y$ direction with respect to a reference plane. This procedure aimed to minimize the difference between the reconstructed track position at the MIMOSA plane and the measured hit position in the same plane. The position resolution was taken as the resolution of the fit performed during the alignment procedure. 

\begin{figure}[h]
\centering
\includegraphics[width=0.9\linewidth]{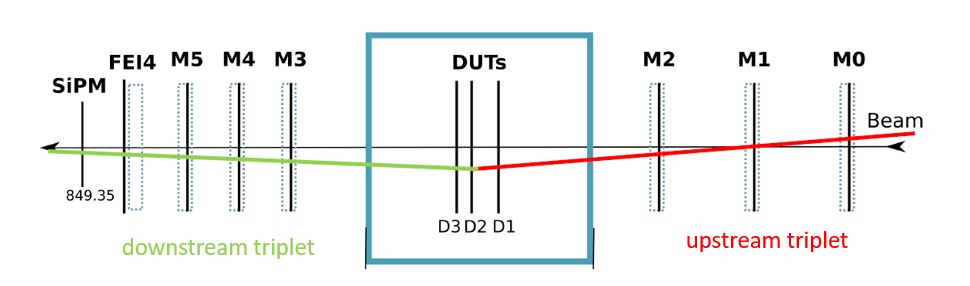}
\caption{
Track reconstruction for data collected during beam tests at DESY. 3D-proto-tracks are reconstructed from the three MIMOSA planes 
before the DUTs (``upstream triplets'') and other 3D-proto-tracks are reconstructed from the three MIMOSA planes after the DUTs 
(``downstream triplets''). The downstream triplets must coincide with a hit in the FE-I4 plane. A complete track is considered to be found if a downstream triplet matches an upstream triplet in the central region with a minimal distance of approach of 150~$\mu$m. }
\label{fig:tracking_desy}
\end{figure}

Given the $z$-position of the MIMOSA planes along the beam axis and the $x$- and $y$-positions of the hits in these planes, three-dimensional (3D) tracks were built. The procedure is described in~\cite{hgtd_tb_paper} and~\cite{hgtd_tb_paper_2018} for data taken during beam tests at CERN SPS with 120~GeV pion beam for which tracks are straightforward, as multiple scattering can be considered negligible. For data collected at DESY with a 5~GeV electron beam, the procedure was slightly different because electrons may be scattered in the equipment around the DUTs. As shown in figure~\ref{fig:tracking_desy}, 3D-proto-tracks are reconstructed from the three MIMOSA planes before the DUTs on the beam trajectory (``upstream triplets''), and other 3D-proto-tracks are reconstructed from the three MIMOSA planes after the DUTs (``downstream triplets''). The downstream triplets must coincide with a hit in the FE-I4 plane. 

For each event, all possible downstream and upstream triplets are reconstructed and a complete track is considered to be found if a downstream triplet matches an upstream triplet in the central region with a minimal distance of approach of 150~$\mu$m. Only events with a single complete track through the six MIMOSA planes are considered. 

\subsection{Track reconstruction for data taken during CERN SPS beam tests with MALTA telescope}
\label{sec:tracking_MALTA}

The event reconstruction, including coarse alignment, fine alignment and tracking, was done using the software package Proteus~\cite{proteus}. Adjacent pixel hits were combined into clusters. The tracks reconstructed from these three telescope layers was extrapolated to the plane of the DUT, taking into account multiple scattering using the General Broken Lines (GBL) formalism. The precision of the positional resolution using six MALTA chips is less than \SI{10}{\micro\metre}.

\subsection{Oscilloscope data processing}
\label{sec:oscilloscope}

A complete description of the LGAD waveform processing can be found in~\cite{hgtd_tb_paper_2018}. In summary, the procedure is the following: after the start and stop points of the pulse were determined, the pedestal was computed in a region outside the pulse where no signal was expected. This pedestal was subtracted from all the data points of the waveform on an event-by-event basis. After that subtraction, the start and stop points were re-computed. Several waveform properties can be extracted at this step such as the maximum of the pulse amplitude, the charge, the rise time, the jitter, the signal-to-noise ratio and the time of arrival (TOA). 

The maximum of the pulse amplitude was estimated as the sample with the maximum amplitude. The collected charge was defined as the integral of the pulse in the signal region divided by the trans-impedance of the read-out board and the gain of the voltage amplifier. The TOA was calculated with a constant fraction discriminator (CFD) method and it was defined as the point at which the signal crosses a predefined fraction ($f_{CFD}$) of its total amplitude. For the analysis of the DESY beam test data, the TOA value at $f_{CFD}=20\%$ was used for the SiPM whereas for the DUTs, the TOA value was chosen to be $f_{CFD}=50\%$. \\ 

The last step was to produce a merged file containing the oscilloscope and telescope data together. In this way, for each event and for each DUT, both the information extracted from the oscilloscope waveforms and the position of the hit of the beam particle in the sensor plane are available for a common analysis.

\section{Sensors performance results}
\label{sec:sec_results}
The studies presented in this paper aim to evaluate the performance of carbon-enriched LGAD sensors at two irradiation levels, 1.5 and 2.5$\times$10$^{15}$~n$_{eq}$/cm$^2$, with particle beams using the reconstructed position of the tracks in the sensor planes. The following LGAD properties have been investigated: the collected charge, the time resolution and the hit reconstruction efficiency. 

\subsection{Data selection}
\label{sec:DataCut}

In all subsequent analyses, background events were removed with a timing cut using a \SI{2}{\nano\second} window from the maximum point of the distribution of the difference between the TOA of the DUT and that of the SiPM. Additionally, the signal amplitude in the SiPM was required to be three times higher than the noise level.

The value of the discriminator of the future ALTIROC chip corresponds to a collected charge of \SI{2}{\femto\coulomb}, as explained in section~\ref{sec:intro}. The efficiency and the time resolution were computed using this threshold in the analysis. 

To measure the global efficiency and the collected charge of the sensors, another condition was added: only the fiducial region of 0.5x0.5 mm2 in the center of the DUTs was used. 
 
\subsection{Collected charge}
\label{sec:collectedcharge}
\begin{figure}[h] 
\centering 
\includegraphics[width=0.5\linewidth]{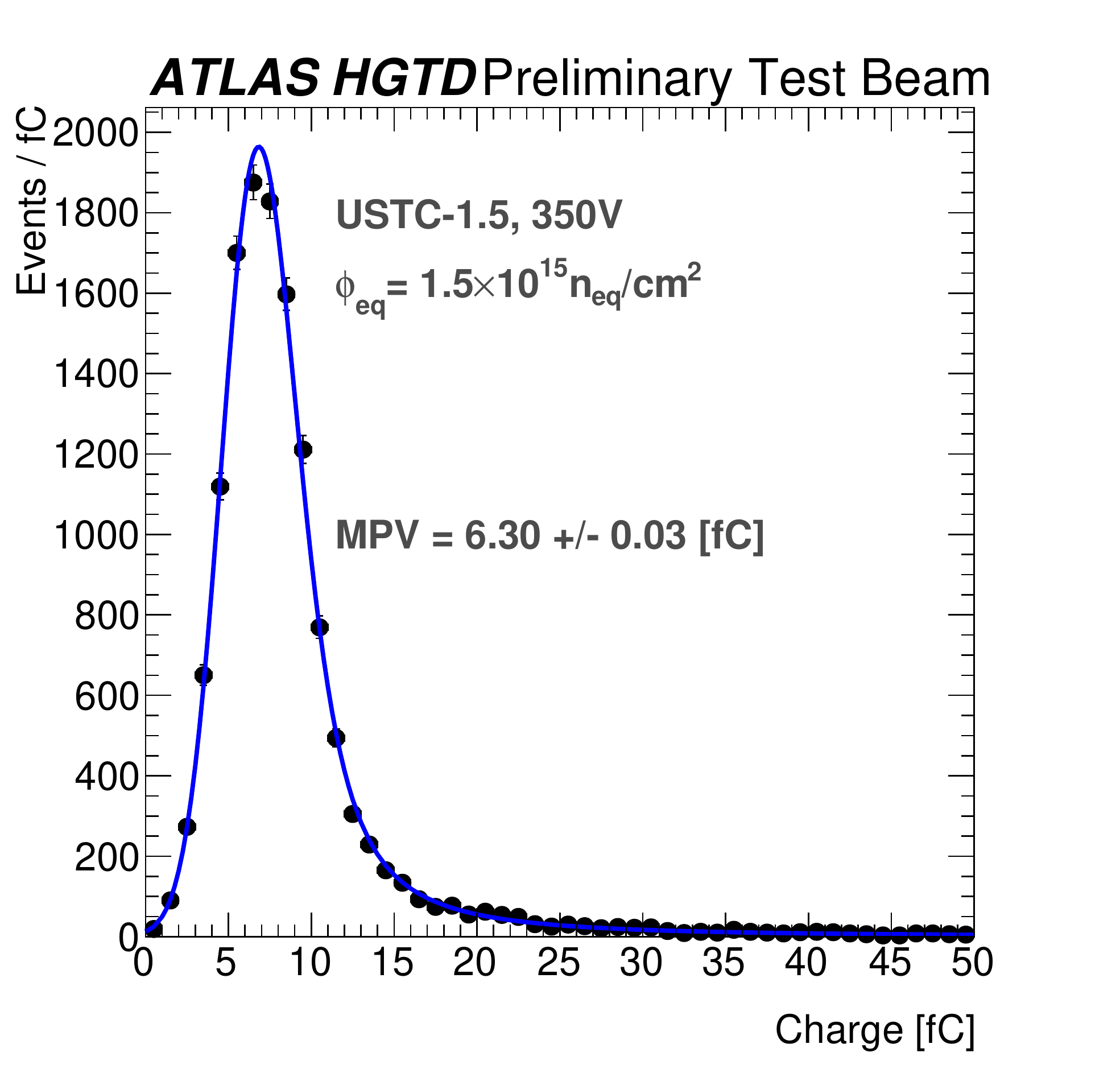} 
\caption{Charge distribution for the sensor USTC-1.5 operated at \SI{350}{\V}. The distribution was fitted with a Landau-Gaussian convoluted function. The collected charge, defined as the most probable value, is $6.3~\pm~0.03$~fC. 
} 
\label{fig:fit_charge} 
\end{figure}

After the background removal, for each DUT, the distribution of the charge (defined in section~\ref{sec:oscilloscope}) was fitted with a Landau-Gaussian convoluted function. The collected charge reported for each sensor was defined as the most probable value (MPV) from this fit. Figure~\ref{fig:fit_charge} shows an example distribution for sensor USTC-1.5. 

\begin{figure}[h]  
\centering
\includegraphics[width=0.7\linewidth]{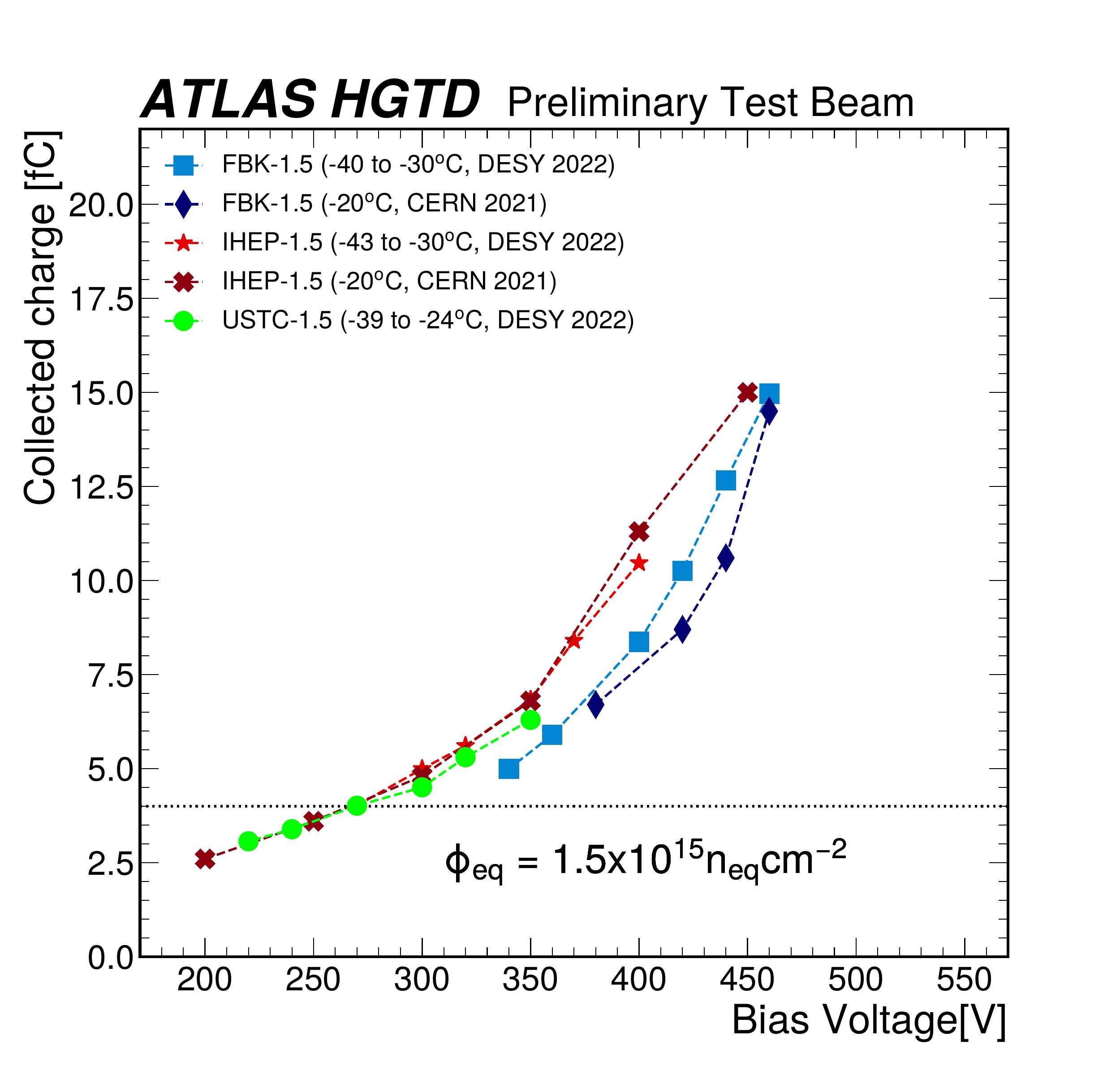} 
\caption{Collected charge as a function of bias voltage for different single-pad sensors: FBK-1.5, USTC-1.5 and IHEP-1.5. Sensors FBK-1.5 and IHEP-1.5 were tested both at DESY with a \SI{5}{\giga\electronvolt} electron beam and at CERN with a \SI{120}{\giga\electronvolt} pion beam, whereas sensor USTC-1.5 was tested only at DESY. The dashed line corresponds to the minimum required charge of \SI{4}{\femto\coulomb} for a good timing measurement with the future HGTD.
}  
\label{fig:ChargevsBV-1.5}  
\end{figure}

\begin{figure}[h]  
\centering
\includegraphics[width=0.7\linewidth]{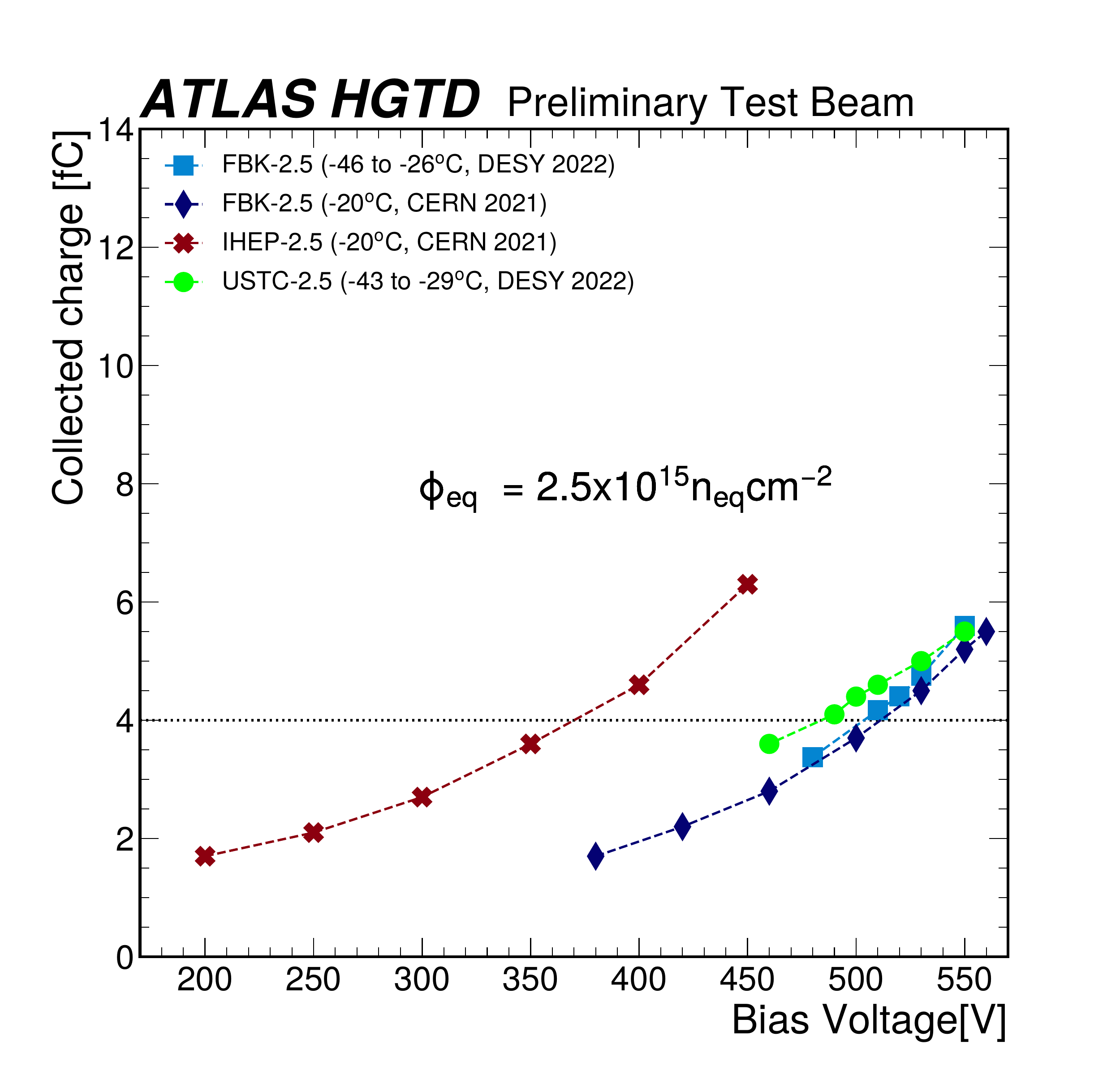}
\caption{Collected charge as a function of bias voltage for different single-pad sensors: FBK-2.5, USTC-2.5 and IHEP-2.5. Sensor FBK-2.5 was tested both at DESY with a \SI{5}{\giga\electronvolt} electron beam and at CERN with a \SI{120}{\giga\electronvolt} pion beam, whereas sensors USTC-2.5 and IHEP-2.5 were tested only at DESY. The dashed line corresponds to the minimum required charge of \SI{4}{\femto\coulomb} for a good timing measurement with the future HGTD.} 
\label{fig:ChargevsBV-2.5} 
\end{figure}

Figures~\ref{fig:ChargevsBV-1.5} and~\ref{fig:ChargevsBV-2.5} show the collected charge as a function of bias voltage for different single-pad sensors irradiated at fluences of $1.5\times10^{15}$~n$_{eq}$/cm$^{2}$ and $2.5\times10^{15}$~n$_{eq}$/cm$^{2}$, respectively. Bias voltages were kept lower than the single event burnout (SEB) voltage at which the sensor dies due to a large amount of energy deposited by a highly energetic particle in the active zone~\cite{seb}. All sensors characteristics are given in section~\ref{sec:sec_sensors}. In both figures, the results come either from sensors tested at DESY with \SI{5}{\giga\electronvolt} electrons or from sensors tested at CERN with \SI{120}{\giga\electronvolt} pions. The methods to analyse these two types of data from different tracking devices are described in sections~\ref{sec:tracking_Patrack} and~\ref{sec:tracking_MALTA}.

Figure~\ref{fig:ChargevsBV-1.5} shows that after an irradiation at a fluence of $1.5\times10^{15}$~n$_{eq}$/cm$^{2}$, the collected charge is above the minimum required charge of \SI{4}{\femto\coulomb} needed for a good timing measurement with the future HGTD detector. This minimum is reached at around \SI{270}{\V} for all sensors, however the sensor FBK-1.5 was tested at higher bias voltages. 
For sensor IHEP-1.5, the same performances were found whether measurements were performed with a \SI{5}{\giga\electronvolt} electron beam at DESY or with a \SI{120}{\giga\electronvolt} pion beam at CERN. For sensor FBK-1.5, at the same bias voltage, the collected charge was slightly higher for measurements at DESY than for those at CERN. Between \SI{350}{\V} and \SI{400}{\V}, which is the range with common bias voltage, the sensors manufactured by IHEP and USTC showed better performances than those from FBK.

The comparison between figures~\ref{fig:ChargevsBV-1.5} and~\ref{fig:ChargevsBV-2.5} shows that the higher the fluence, the worse the collected charge at the same bias voltage. Indeed, figure~\ref{fig:ChargevsBV-2.5} shows that after an irradiation at a fluence of $2.5\times10^{15}$~n$_{eq}$/cm$^{2}$, the minimum required collected charge of \SI{4}{\femto\coulomb} is reached for a bias voltage around \SI{370}{\V} for sensor IHEP-2.5, around \SI{500}{\V} for sensor FBK-1.5 and \SI{470}{\V} for sensor USTC-2.5. Comparing the different sensors at this fluence, those manufactured by IHEP show better performance than sensors from FBK and USTC at the same bias voltage. This discrepancy is less significant at a fluence of $1.5\times10^{15}$~n$_{eq}$/cm$^{2}$.

\subsection{Time resolution}
\label{sec:timereso}
The time resolution is one of the key parameters when assessing LGAD sensor performance and has been explained in detail in~\cite{hgtd_tb_paper_2018}. To extract the DUTs' time resolutions, the distributions of the difference between the TOA of the DUTs and that of the time reference device (SiPM at DESY and CNM-0 at CERN) was used. 

%
%
 
During the DESY beam test campaign, there were two DUTs (labelled as $1$ and $2$ in equation~\ref{eq:timeres}) and one SiPM connected to the oscilloscope. Hence, it was possible to compute three time differences on an event-by-event basis between all these devices: 
\begin{equation}
\left \{
\begin{array}{c}
t_1 - t_2\\
t_1 - t_{SiPM}\\
t_2 - t_{SiPM}\\
\end{array}
\right.
\label{eq:timeres}
\end{equation}

The three distributions were fitted with a gaussian function, each of them giving a width $\sigma_{ij}$ where $i$ and $j$ are the names of the devices used for each distribution. Assuming that the time resolution of the devices are independent, each of measured resolution $\sigma_{ij}$ corresponds to 
\begin{equation}
\sigma_{ij}=\sigma_i \oplus \sigma_j
\end{equation}
where $\sigma_i$ and $\sigma_j$ are the resolutions of each device. This gives a system of three equations and three unknowns which can be solved analytically.  

No matter which DUTs are tested, the SiPM time resolution should have the same value. Overall, during the DESY beam test campaign, the SiPM time resolution was measured 17 times with this method and its average resolution was found to be $\sigma_{SiPM}=62.6 \pm 0.6$~ps with variations up to 
7~\% due to differences in running conditions (e.g. bias voltage settings).  

During the CERN beam test campaign, the time reference was given by CNM-0 with a known time resolution of 54.8~ps. The time resolution of the DUT was computed as for the DESY analysis: the distribution of the difference between the TOAs of the reference LGAD and the DUT was fitted with a gaussian function, giving a width $\sigma_{\text{DUT-LGAD}}$. The time resolution of the DUT is then directly given by the equation:  
\begin{equation}
\sigma_{\text{DUT-ref~LGAD}}=\sigma_{\text{DUT}} \oplus \sigma_{\text{ref~LGAD}}
\end{equation}
 
where $\sigma_{\text{ref~LGAD}}~=~54.8$~ps. 

\begin{figure}[h]   
\centering 
\includegraphics[width=0.7\linewidth]{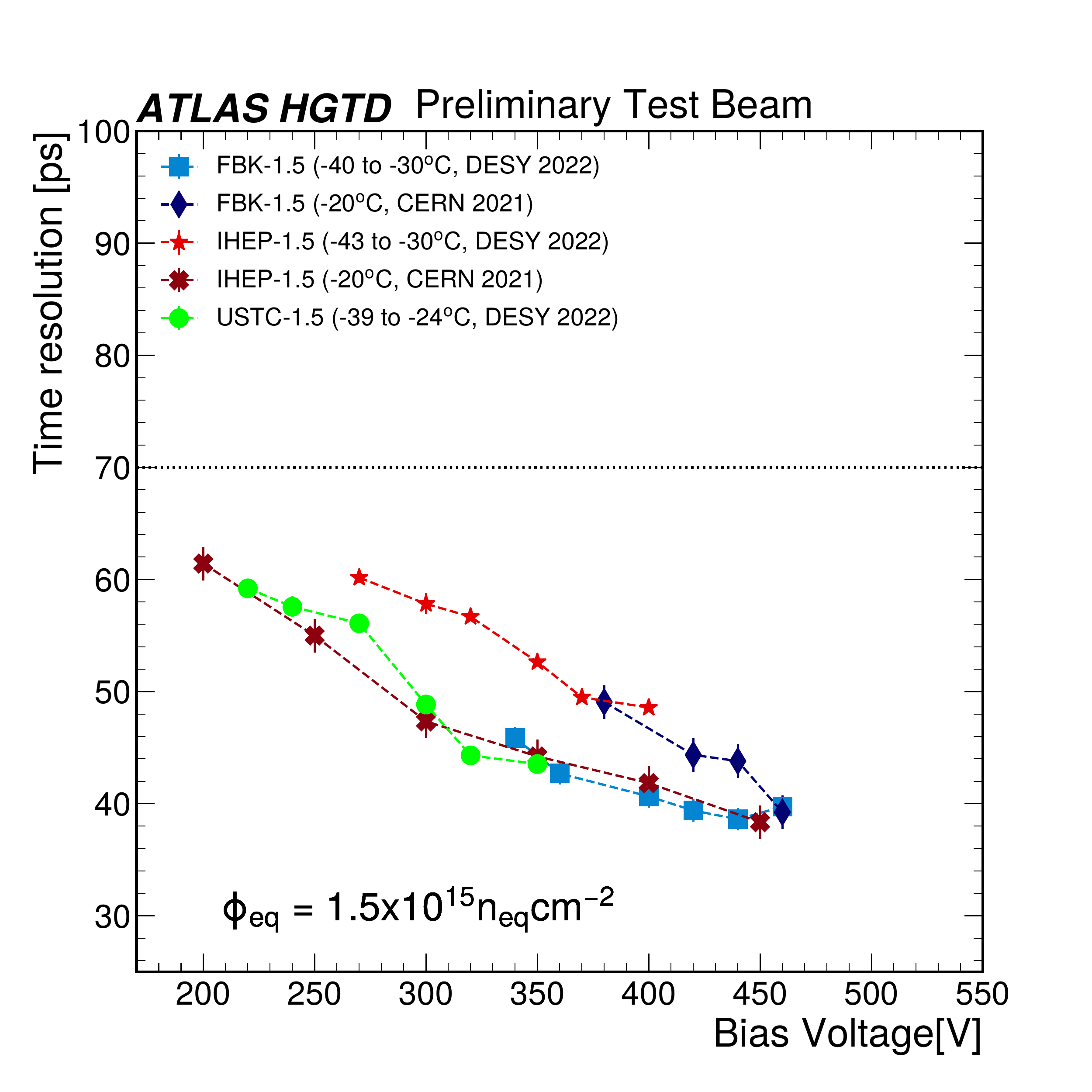}  
\caption{Time resolution as a function of bias voltage for different single-pad sensors: FBK-1.5, USTC-1.5 and IHEP-1.5. The measurements were performed in the conditions described in figure~\ref{fig:ChargevsBV-1.5}. The dashed line at \SI{70}{\pico\second} represents the maximum time resolution permitted by the future HGTD after irradiation.}
\label{fig:TimeResvsBV-1.5}   
\end{figure}

\begin{figure}[h]   
\centering 
\includegraphics[width=0.7\linewidth]{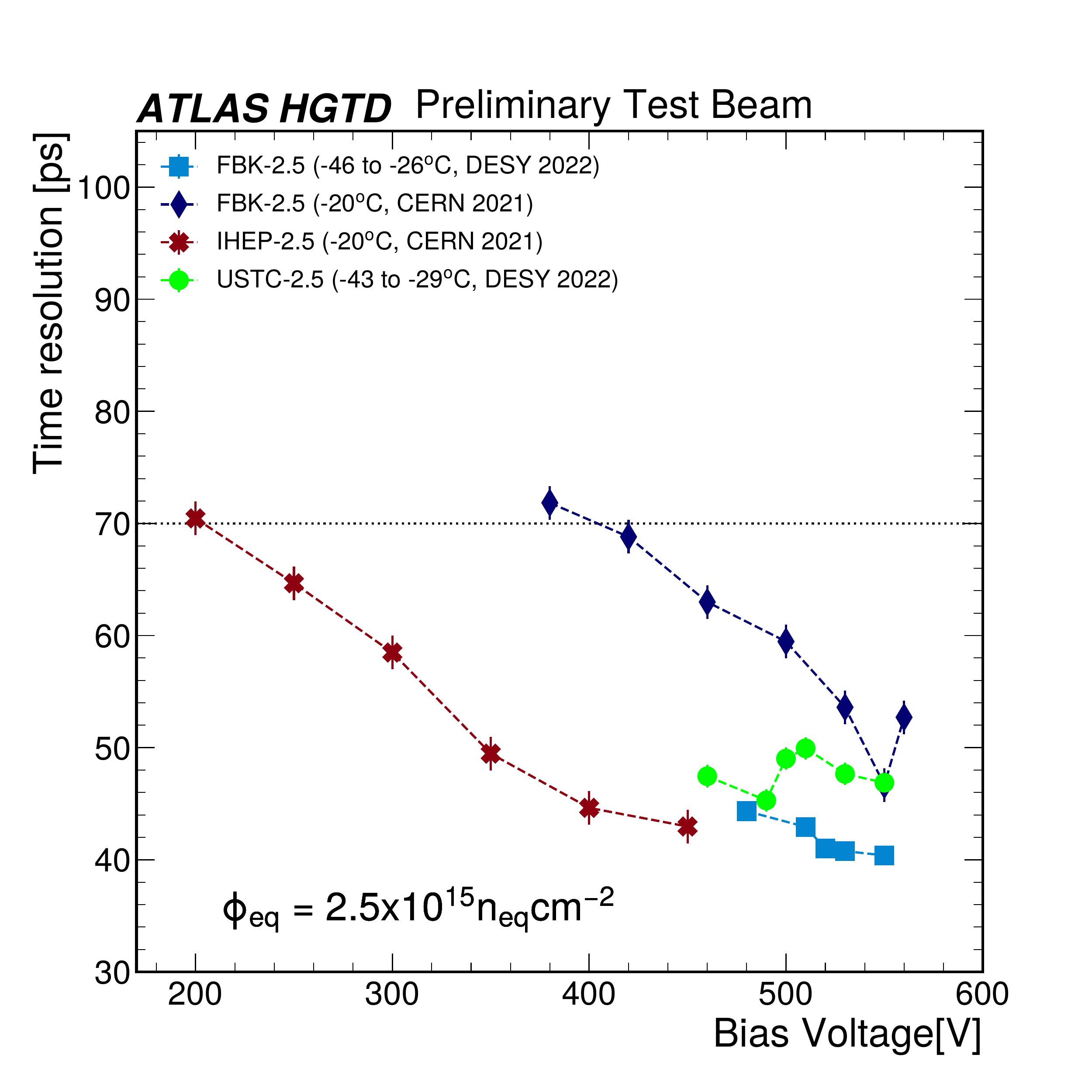} 
\caption{Time resolution as a function of bias voltage for different single-pad sensors: FBK-2.5, USTC-2.5 and IHEP-2.5. The measurement were performed in the conditions described in figure~\ref{fig:ChargevsBV-2.5}. The dashed line at \SI{70}{\pico\second} represents the maximum time resolution permitted by the future HGTD after irradiation.}
\label{fig:TimeResvsBV-2.5}  
\end{figure}

Figures~\ref{fig:TimeResvsBV-1.5} and~\ref{fig:TimeResvsBV-2.5} show the time resolution as function of the bias voltage for different single-pad sensors irradiated at fluences of $1.5\times10^{15}$~n$_{eq}$/cm$^{2}$ and $2.5\times10^{15}$~n$_{eq}$/cm$^{2}$ respectively. In both figures, the results come either from sensors tested at DESY with \SI{5}{\giga\electronvolt} electrons or from sensors tested at CERN with \SI{120}{\giga\electronvolt} pions. 

Both figures show an improvement in the time resolution with higher bias voltage, as expected. Figure~\ref{fig:TimeResvsBV-1.5} shows that sensors irradiated at a fluence of $1.5\times10^{15}$~n$_{eq}$/cm$^{2}$ can achieve a time resolution of around \SI{39}{\pico\second} for sensors FBK-1.5 and IHEP-1.5 at a bias voltage of 450~V and of \SI{44}{\pico\second} for sensor USTC-1.5 at a bias voltage of 350~V. 

Figure~\ref{fig:TimeResvsBV-2.5} shows that even after an irradiation at a fluence of $2.5\times10^{15}$~n$_{eq}$/cm$^{2}$, time resolution measurements can almost reach the same values as after an irradiation at a lower fluence of $1.5\times10^{15}$~n$_{eq}$/cm$^{2}$. Indeed, at a higher bias voltage of \SI{550}{\V}, sensors FBK-2.5 and USTC-2.5 can achieve a time resolution of \SI{40}{\pico\second} and of \SI{46}{\pico\second}, respectively. Sensor IHEP-2.5 achieves a time resolution of around \SI{43}{\pico\second} at the same bias voltage of 450V as for the measurement at a fluence of $1.5\times10^{15}$~n$_{eq}$/cm$^{2}$.

\subsection{Hit reconstruction efficiency}
\label{sec:hiteff}
The hit reconstruction efficiency is defined as the number of reconstructed tracks giving a signal on the center of the sensor for which the charge in the sensor is greater than a given threshold value, Q$_{cut}$, divided by the total number of reconstructed tracks crossing the same fiducial region:  
\begin{equation} 
\text{Hit Efficiency}=\frac{\text{Reconstructed tracks with } q> Q_{cut}}{\text{Total reconstructed tracks}} 
\end{equation} 

As explained in the paragraph~\ref{sec:DataCut}, Q$_{cut}$ is set to \SI{2}{\femto\coulomb}.

\subsubsection{Efficiency as a function of the bias voltage}

\begin{figure}[h]    
\centering  
\includegraphics[width=0.7\linewidth]{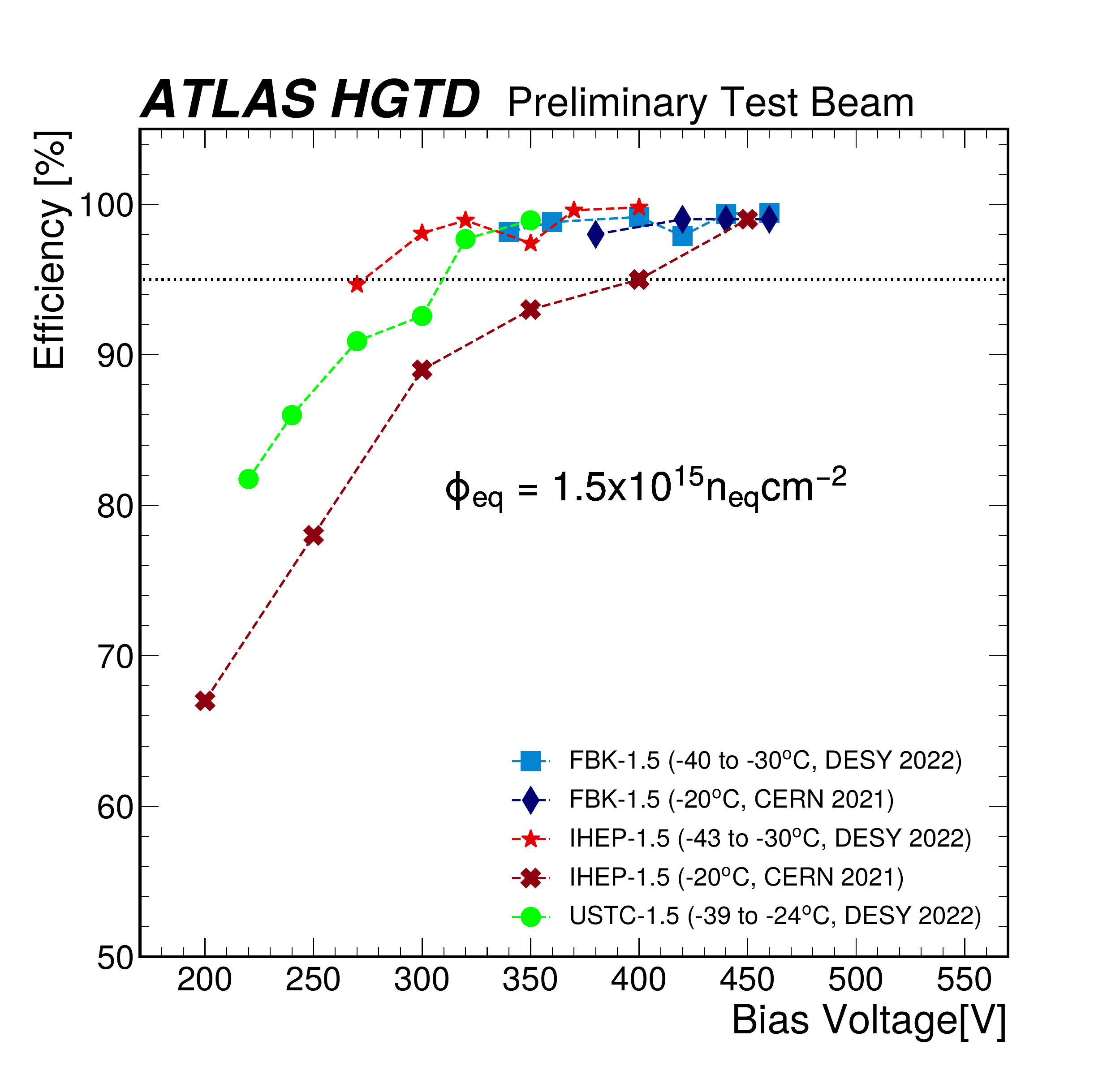}
\caption{Efficiency as a function of bias voltage for different single-pad sensors: FBK-1.5, USTC-1.5 and IHEP-1.5. The measurements were performed in the conditions described in figure~\ref{fig:ChargevsBV-1.5}. The dashed line at 95\% corresponds to the efficiency needed for required operation of the future HGTD after irradiation.} 
\label{fig:EfficvsBV-1.5}    
\end{figure}

\begin{figure}[h]    
\centering  
\includegraphics[width=0.7\linewidth]{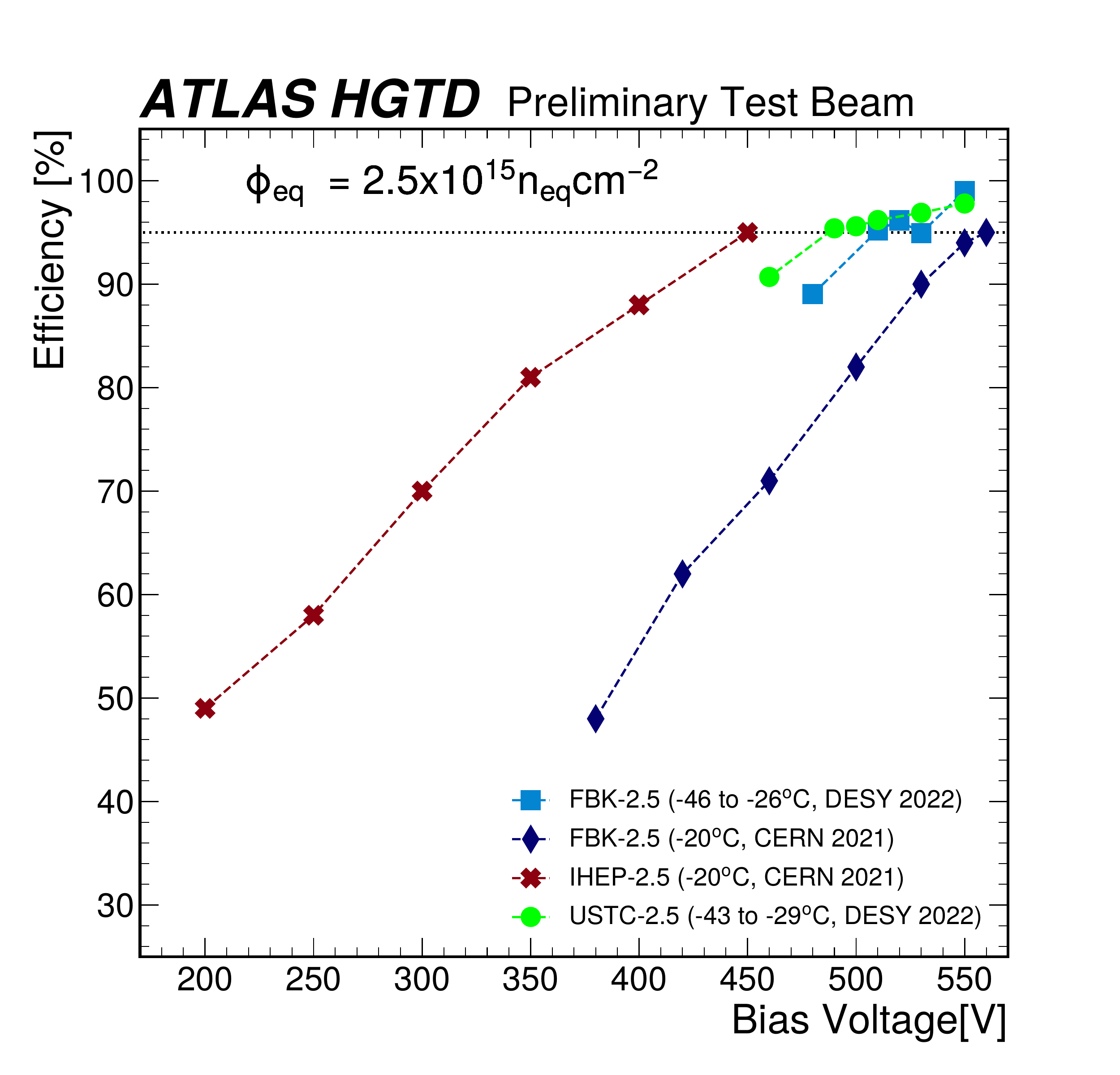}
\caption{Efficiency as a function of bias voltage for different single-pad sensors: FBK-2.5, USTC-2.5 and IHEP-2.5. The measurements were performed in the conditions described in figure~\ref{fig:ChargevsBV-2.5}. The dashed line at 95\% corresponds to the efficiency needed for required operation of the future HGTD after irradiation.}
\label{fig:EfficvsBV-2.5}   
\end{figure}

Figures~\ref{fig:EfficvsBV-1.5} and~\ref{fig:EfficvsBV-2.5} show the efficiency as a function of bias voltage for different single-pad sensors irradiated at fluences of $1.5\times10^{15}$~n$_{eq}$/cm$^{2}$ and $2.5\times10^{15}$~n$_{eq}$/cm$^{2}$, respectively. In both figures, the results come either from sensors tested at DESY with \SI{5}{\giga\electronvolt} electrons or from sensors tested at CERN with \SI{120}{\giga\electronvolt} pions.  
  
Both figures show that a higher efficiency can be reached by increasing the bias voltage. Figure~\ref{fig:EfficvsBV-1.5} shows that all DUTs irradiated at a fluence of $1.5\times10^{15}$~n$_{eq}$/cm$^{2}$ can achieve an efficiency of more than 98.9\%. Figure~\ref{fig:EfficvsBV-2.5} shows that at a higher fluence of $2.5\times10^{15}$~n$_{eq}$/cm$^{2}$, the bias voltage needs 
to be higher than at  $1.5\times10^{15}$~n$_{eq}$/cm$^{2}$ to reach almost similar performances or just below. Indeed, at a fluence of $2.5\times10^{15}$~n$_{eq}$/cm$^{2}$, the sensors tested at DESY reach an efficiency of 99\% for FBK-2.5 and 98.1\% for USTC-2.5. This is well beyond the efficiency of 95\% required for good operation of the future HGTD after irradiation. As for the time resolution, the effects of the temperature can also be seen in the efficiency: the sensors tested at CERN at a higher temperature of \SI{-20}{\degree C} reach a lower efficiency of 95\%. 

\subsubsection{Efficiency uniformity}

\begin{figure}[htbp]
\centering
\begin{tabular}{cc}
\includegraphics[width=0.5\linewidth]{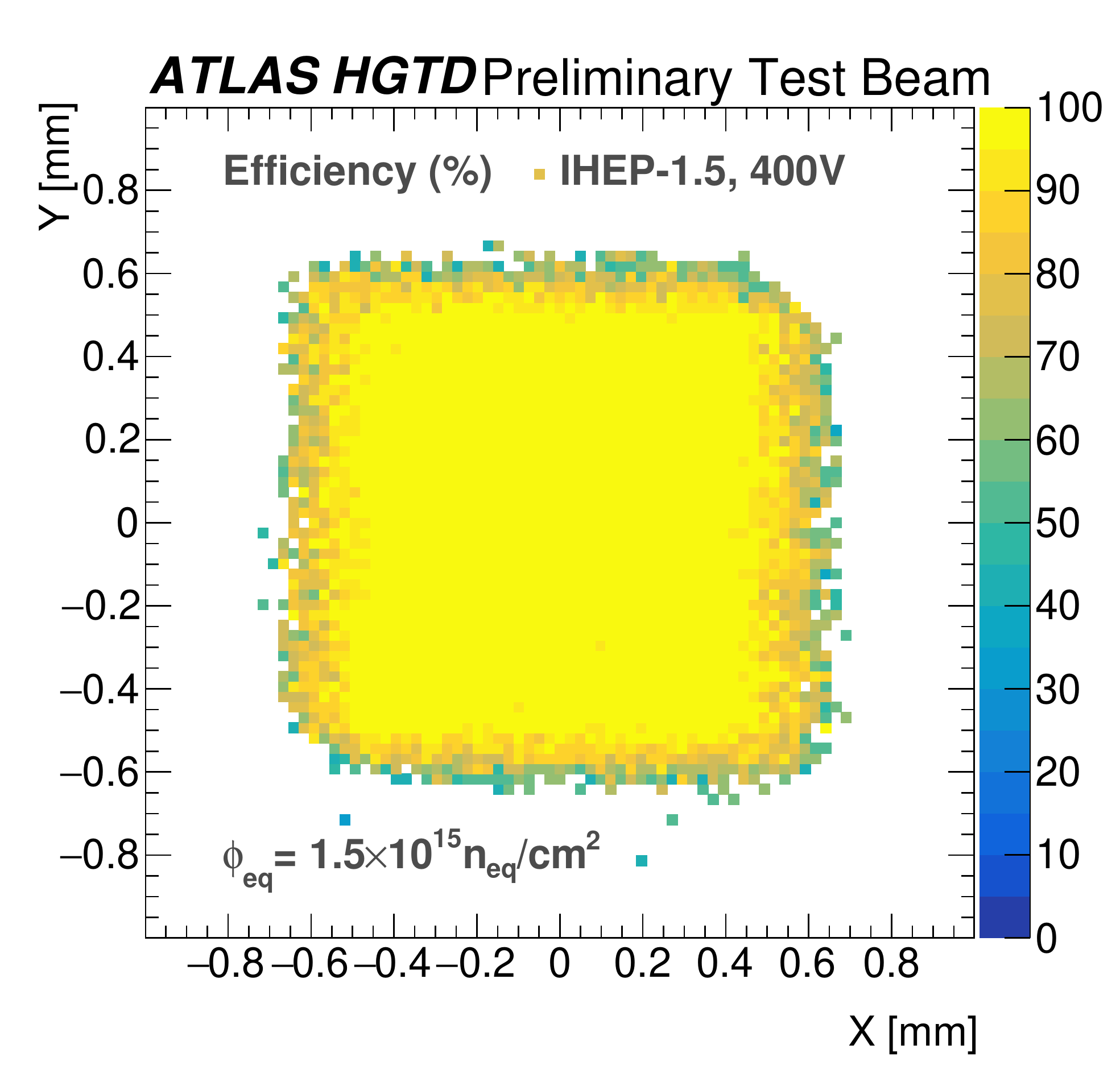} &
\includegraphics[width=0.5\linewidth]{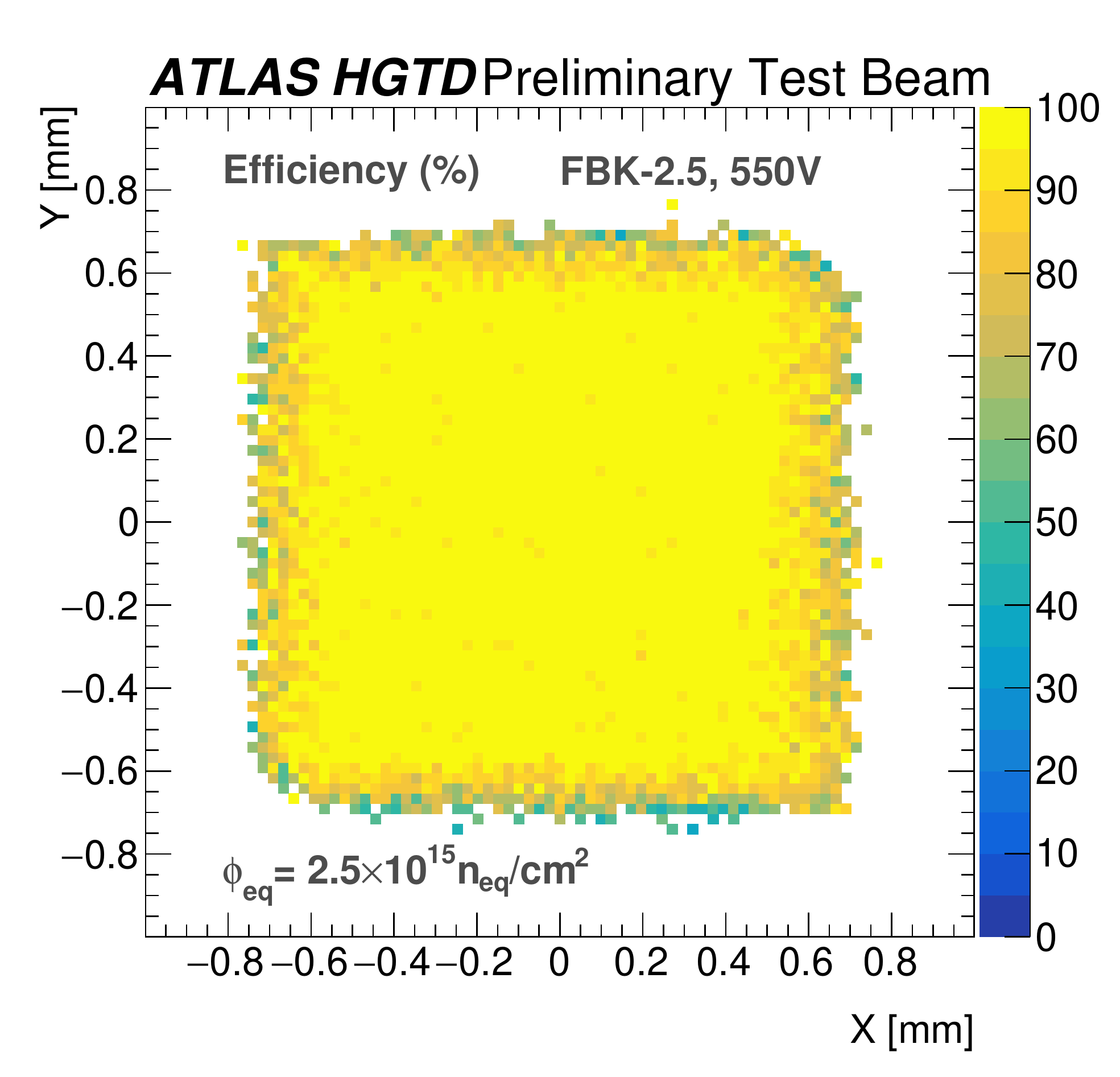} \\
(a) & (b) \\
\end{tabular}
\caption{2D maps of the efficiency as a function of hit position in the sensor plane. Figure (a) was made for the sensor IHEP-1.5 operated at a bias voltage of \SI{400}{\V}. Figure (b) was made for the sensor FBK-2.5 operated at a bias voltage of \SI{550}{\V}. Both sensors were tested at DESY with a \SI{5}{\giga\electronvolt} electron beam.}
\label{fig:eff2D}
\end{figure}

Figures~\ref{fig:eff2D}~(a) and~\ref{fig:eff2D}~(b) show the two-dimesional (2D) map of the efficiency as a function of the hit position for the IHEP-1.5 sensor (figure (a)) and the FBK-2.5 sensor (figure (b)). Both sensors were tested at DESY with a 5~GeV electron beam. They were enclosed in a styrofoam box, filled with dry ice, which was moving upwards during data taking due to the evaporation of the dry ice. Movements, temperature variation, and other factors may contribute to the smeared figures. 

\begin{figure}[htbp]
\centering
\includegraphics[width=0.7\linewidth]{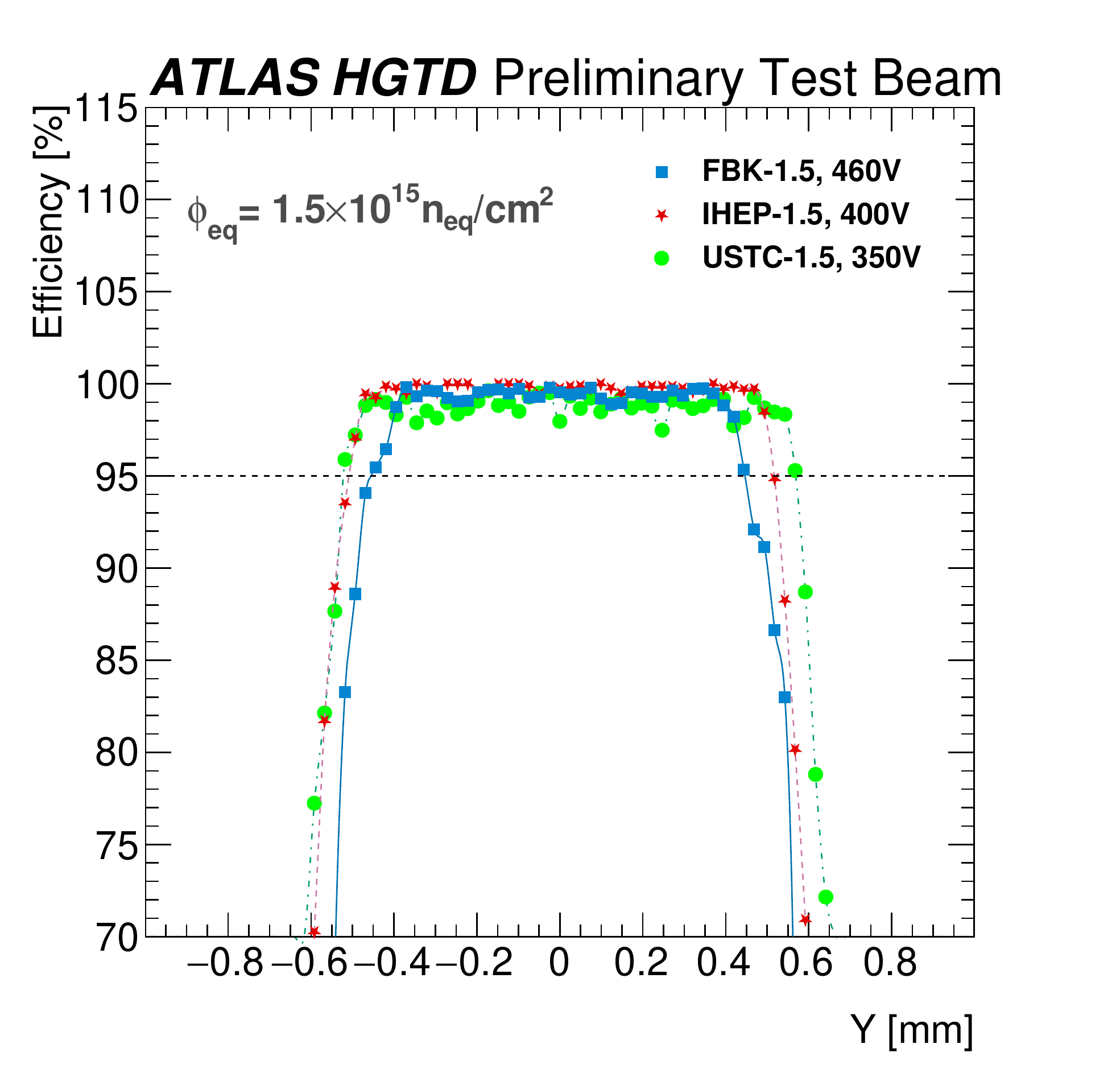} 
\caption{Projections on the $y$-axis of the efficiency for sensors irradiated at a fluence of $1.5\times10^{15}$~n$_{eq}$/cm$^{2}$ and tested at DESY with a 5~GeV electron beam. The dashed line at 95\% corresponds to the efficiency needed for required operation of the future HGTD after irradiation.}
\label{fig:eff_profile}
\end{figure}

For more details, figure~\ref{fig:eff_profile} shows the uniformity of the efficiency along the y-axis for the sensors irradiated at a fluence of $1.5\times10^{15}$~n$_{eq}$/cm$^{2}$ and tested at DESY with a \SI{5}{\giga\electronvolt} electron beam. The efficiency of the sensor USTC-1.5 shows some variation up to 1.5\% of the mean efficiency on the plateau. Figure~\ref{fig:eff_profile} also allows to extract the size of the region where the efficiency was larger than 99.8\% of the efficiency computed in the center of the DUT. These sizes are given in table~\ref{tab:sizeplateau} for the three sensors. Table~\ref{tab:sizeplateau} also gives the size of the region where the efficiency is larger than 95\%, which is the required efficiency for irradiated sensors. These numbers can be compared to the nominal surface of 1.3$\times$1.3~$mm^2$ of the DUTs. 

\begin{table}[htbp]
\centering
\caption{Mean efficiency in the center of each DUT. The bias voltage applied for each measurement is also indicated. 
The 5th column contains the size of the region where the efficiency is larger than 99.8\% of the mean efficiency given in column 4. 
The last column contains the size of the region where the efficiency is larger than 95\%, which is the minimal efficiency needed for required operation of the future HGTD after irradiation.} 
\begin{tabular}{|c|c|c|c|c|c|}
\hline
\textbf{Sensors} & \textbf{Bias Voltage} & \textbf{Temperature} & \textbf{Efficiency}       & \multicolumn{2}{c|}{\textbf{Size of the plateau} }  \\   
        &   \textbf{[V]}     &  \textbf{[\SI{}{\degree C}]} & \textbf{in the DUT center} &  \textbf{at 99.8\% of } & \textbf{if eff > 95\% } \\
          &      &    & \textbf{[\%]}   &  \textbf{DUT eff [$\mu$m]} & \textbf{ [$\mu$m]} \\     
\hline  
FBK-1.5 & 460 & -32 & 99.4 & 740  & 863 \\
USTC-1.5 & 350 & -33 & 98.9 & 937  & 1085 \\
IHEP-1.5 & 400 & -38 & 99.8 & 888 & 987 \\
\hline
\end{tabular}
\label{tab:sizeplateau}
\end{table} 

\FloatBarrier

\section{Conclusion and outlook}
\label{sec:conclusion}
The High Granularity Timing Detector requires high-performance and radiation-resistant sensors. LGADs have been studied both in terms of radiation resistance and performance. Beam test campaigns have led us to choose promising technologies studied in this paper. Carbon-enriched LGAD samples from three vendors were considered: FBK, IHEP-IME and USTC-IME. The LGADs were irradiated to simulate their end-of-life state and studied under particle beams at DESY and CERN in 2021 and 2022.

Although irradiated at fluences of $1.5$ to $2.5 \times 10^{15}$~n$_{eq}$/cm$^{2}$, the LGADs presented in this paper were operated at voltages below \SI{550}{\V}. Under these conditions, they achieved the objectives of a collected charge of more than \SI{4}{\femto\coulomb} while guaranteeing an optimum time resolution below \SI{70}{\pico\second}. An efficiency larger than 95\% uniformly over the sensor's surface is obtained with a charge threshold of \SI{2}{\femto\coulomb}. These results confirm the feasibility of an LGAD-based timing detector for HL-LHC.

\section*{Acknowledgements}

The authors gratefully acknowledge CERN and the SPS staff for successfully operating the North Experimental Area and for continuous supports to the users.
Measurements leading to the SPS results have been carried out using a MALTA beam telescope as part of the CERN ATLAS R\&D programme. The authors thank MALTA team for their successful running of the telescope during the data-taking.
The measurements leading to these results have been performed at the Test Beam Facility at DESY Hamburg (Germany), a member of the Helmholtz Association (HGF). The authors gratefully acknowledge DESY staff for their support to the users.
This work was partially funded by MINECO, Spanish Government, under grant RTI2018-094906-B-C21. This project has received funding from the European Union's Horizon 2020 research and innovation programme under the Marie Sklodowska-Curie grant agreement No. 754510. This work was partially funded by: the Spanish Government, under grant FPA2015-69260-C3-2-R and SEV-2012-0234 (Severo Ochoa excellence programme); and by the H2020 project AIDA-2020, GA no. 654168.
This work is partially supported by “the Fundamental Research Funds for the Central Universities” of China (grant WK2030040100), the National Natural Science Foundation of China (No. 11961141014) and the Chinese Academy of Sciences (Grant NO. : GJJSTD20200008).

\pagebreak




\end{document}